\documentclass[conference,compsoc]{IEEEtran}
\ifCLASSOPTIONcompsoc
  \usepackage[nocompress]{cite}
\else
  \usepackage{cite}
\fi
\usepackage{graphicx}

\ifCLASSINFOpdf
\else
\fi
\usepackage{amsfonts} 
\usepackage{amsmath}
\usepackage{amsthm}
\usepackage{dsfont}
\newtheorem{theorem}{Theorem}
\newtheorem{prop}{Proposition}
\usepackage{caption}
\usepackage[T1]{fontenc}
\usepackage[rgb]{xcolor}
\usepackage{smartdiagram}

\begin{document}
%
\title{Salt-based autopeering for DLT-networks}

\author{\IEEEauthorblockN{Sebastian Müller\IEEEauthorrefmark{1}\IEEEauthorrefmark{2},
Angelo Capossele\IEEEauthorrefmark{2},
Bartosz Ku\'{s}mierz\IEEEauthorrefmark{2},
Vivian Lin\IEEEauthorrefmark{2},\\
Hans Moog\IEEEauthorrefmark{2},
Andreas Penzkofer\IEEEauthorrefmark{2},
Olivia Saa\IEEEauthorrefmark{2},
William Sanders\IEEEauthorrefmark{2},
Wolfgang Welz\IEEEauthorrefmark{2}}
\IEEEauthorblockA{\IEEEauthorrefmark{1}Aix Marseille Universit\'e, CNRS, Centrale Marseille, I2M - UMR 7373, \\ 13453 Marseille, France\\ Email: sebastian.muller@univ-amu.fr}
\IEEEauthorblockA{\IEEEauthorrefmark{2}IOTA Foundation\\
Berlin, Germany\\
Email: research@iota.org}}


\maketitle

\begin{abstract}
The security of any Distributed Ledger Technology (DLT) depends on the safety of the network layer. Much effort has been put into understanding the consensus layer of DLTs. However, many network layer designs seem ad-hoc and lack a careful analysis of the influence of the design decisions on the whole DLT system. 
We propose a salt-based automated neighbor selection protocol that shows the inherent tradeoffs of certain design decisions and allows a quantitative treatment of some network topology requirements. This example may serve as a design framework and facilitate future research.  We provide a selection of results from simulations to highlight some tradeoffs in the design decisions.  
\end{abstract}


%
\IEEEpeerreviewmaketitle

\section{Introduction}
Distributed Ledger Technologies (DLTs) constitute a recent approach to recording and sharing data across multiple data stores. 
Fundamental elements in the design of a DLT are the communication level, transaction data storage, and a distributed consensus protocol. Research and industry devoted much attention to consensus algorithms, the so-called core of each DLT. However, most of the works on DLT usually overlook the essential aspect of the underlying communication layer and give very few details on the corresponding peer-to-peer (P2P) network.  

A recent line of work \cite{SecurityPoW, SelfishMining, eclipseOnBitcoin, TamperingDelivery, marcus2018low} clearly shows that the security properties of a DLT rely on the security of its underlying P2P network. We refer to \cite{neudecker2018network} for a survey on the network layer aspects of permissionless DLTs.
With such a diversity of DLTs at hand, a one-size-fits-all design paradigm for an automated neighbor selection mechanism seems far from being effective. For example, Bitcoin and Ethereum rely on flat random graph topologies, but Cardano and Ripple use a more hierarchical approach where nodes may play different roles, which influence how nodes connect.
Only a careful analysis of the specific requirements of the DLT may tell whether the underlying P2P network should enforce a particular network topology or allow the verifiability or hiding of peer connections.

This problem is known to the scientific community; in particular, \cite{neudecker2018network}  and \cite{SOK} constitute a ``call-to-arms'' to the networking community to develop network layers that address the specific needs of DLTs. 
\definecolor{babyblueeyes}{rgb}{0.63, 0.79, 0.95}
\smartdiagramset{
 uniform color list=babyblueeyes for 7 items,
 planet size=2.5cm,
 planet text width=2.5cm,
 satellite size=1.5cm, 
 satellite text width=1.5cm,
 satellite font=\footnotesize,
 distance planet-text=0,
 distance planet-satellite=3.0cm,
 /tikz/connection planet satellite/.append style={->}
 } 
 \begin{figure}[ht]
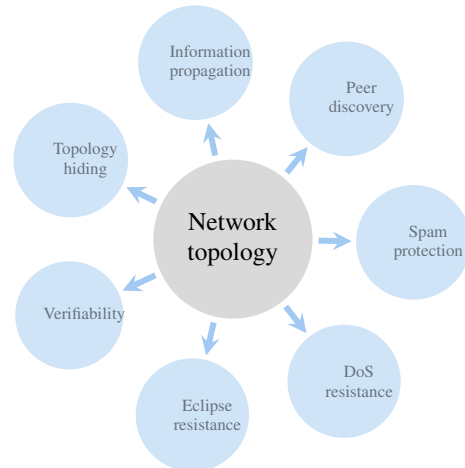

 \begin{center}
 \scalebox{0.8}{
 \smartdiagram[constellation diagram]{
  Network topology,
  Peer discovery,
  Information propagation,
  Topology hiding,
  Verifiability,
  Eclipse resistance,
  DoS resistance,
  Spam protection}}
  \end{center}
  \caption{Main requirements on network topology.}
  \label{fig:requirements}
  \end{figure}
We propose a framework for neighbor selection that allows a systematic analysis of the most important properties for  DLTs. While every DLT may have particular needs on the underlying P2P network, they all should verify some basic requirements. Figure \ref{fig:requirements} gives a high level overview of the requirements on the network topology in DLTs. The main purpose of a P2P is to ensure the propagation of data to as many participants as fast as possible. This is required to allow the DLT to establish which transactions are valid and which are not. 
 Another essential requirement of a P2P network is robustness against eclipse attacks. In an eclipse attack, an attacker controls all of the victim's connections, and hence isolates the victim from the rest of the honest peers. Consequently, an attacker can filter or alter the victim's view on the DLT, and filter or delay messages issued by the victim.  This attack is of particular interest in permissionless systems, where an attacker could create a large number of nodes, a so-called Sybil attack \cite{TheSybilAttack}, to increase the chances to eclipse the victim.

\subsection{Contribution}
We propose a framework for the design of neighbor selection protocols that allows the analytical treatment of inherent tradeoffs of some design decisions. We present a specific example  is based on hash chains. These hash chains allow increasing the robustness against eclipse attacks by maintaining reasonable concealing of the network topology. We also present simulation results on network stability. 

The implementation of the proposed solution is released as an open source project~\cite{simulator} and adds up to this paper's contribution. It has been used to obtain our simulations results and is currently used by the IOTA 2.0 DevNet\footnote{{http://ressims.iota.cafe:28080/autopeering}}.

\section{Related work}
In the DLT space, the DHT is a common approach for peering, and its variations are used by popular second-generation DLTs such as Ethereum and Cardano.  Let us note that Kademlia and its derivatives, e.g., \cite{SKademlia}, have been used for many years; however, there is no analytical proof for its performances and robustness. There have been more theoretical approaches, e.g., \cite{brahms, towardsPrcaticalCommunic, dynamicDC}, but they were not able to establish a practicable alternative. A main focus in the analysis of P2P networks lies in  the vulnerabilities and possible defenses against eclipse attacks, e.g. see \cite{eclipseOnBitcoin}, \cite{marcus2018low}, \cite{castro2002secure}, \cite{hildrum2003asymptotically}, and \cite{singh2006eclipse}.

\section{Autopeering model}
We require every node to have a unique identifier. This identifier does not only allow to globally address and recognize nodes but it may enable an identity-based Sybil protection. Moreover, it allows the implementation of mechanisms like \textit{node accountability} where {nodes can gain or lose reputation or be held responsible for their actions}.
    
The most straightforward approach for implementing this is to automatically generate a new private/public key pair for every new node, where the public key (or its hash) serves as an identifier and the private key can be used by this node to sign messages. The set of all nodes in the network is denoted by $\mathcal{N}$ and there are $N = |\mathcal{N}|$ nodes in total, denoted $n_1,...,n_N$. We call $n_i$ the node id of the $i$th node, e.g., the hash of the public key of the node. 

To be able to choose peers, every node needs a list of potential peering partners. In a permissionless system, this list will not be static but will constantly change, since nodes can leave or join the network at any point in time.  Since we limit our analysis on neighbor selection we refer to \cite{neudecker2018network} for a discussion on peer discovery aspect of the autopeering. 

\subsection{Classifying neighbors according to their roles}
For any autopeering to work in an unreliable environment, a node needs to not only be able to initiate outgoing connections to other nodes but also be willing to accept incoming connection requests. 
In general, the average number of incoming connections needs to match the number of outgoing connections in the network. 
In this paper, we limit ourselves to the case where all nodes have the same number of incoming and outgoing connections. We define two distinct groups of connections or neighbors:
\begin{itemize}
    \item Outbound Neighbors: The neighbors that the node proactively chose from its list.
    \item Inbound Neighbors: The neighbors that chose the node as their peer from their list.
\end{itemize}
As outbound neighbors are chosen proactively, it is generally harder for an adversary to control a node's outbound connections. For instance, a natural strategy to overtake the inbound connections of a victim is to send a high number of requests from different nodes controlled by a single entity. In the next section, we introduce a locally verifiable autopeering that allows better protection of the inbound connections.
\subsection{Locally verifiable autopeering}\label{sec:locally_verifiable}
Neighbor selection can be done purely randomly or following some different criterion. We propose that each node maintains a list of outbound and inbound scores. Outbound connections are chosen to minimize the outbound scores and inbound connections are selected to minimize the inbound scores. Both scores are based on periodically updated salts (i.e., random data used as an additional input to a hash function).  Each node maintains a pair of salts: the private and the public salt. The private salt is used to ``randomize'' the inbound neighbors and the public salt can be used to verify that a requesting node adheres to the protocol.
More specifically, for each node $n_i\in \mathcal{N}$ there exists a public node id $n_i$, a public outbound salt $Salt_{pub}$, and a private inbound salt $Salt_{priv}$.
Node $n_i$ chooses a candidate for the outbound connection based on the outbound scores. We define the outbound score between requesting node $n_i$ and node $n_j$ as
\begin{eqnarray}
     S_{{out}}(n_i,n_j) = 
    hash(n_i || n_j|| Salt_{pub}(n_i)),
\end{eqnarray}
where $||$ stands for the concatenation operator and the hash is normalized to return  values in the interval $[0,1]$.  
To request a new outbound connection a node $n_i$ starts requesting from the smallest to the highest score $S_{{out}}(n_i,\cdot),$ until the maximal number $k$ of outbound neighbors is attained.

A node that is being requested for a connection will base its decision for acceptance on the following inbound score. For  a node  $n_i$ we define the inbound score of $n_j$ as follows:
\begin{eqnarray}
    S_{{in}}(n_i,n_j)  =      
    hash(n_i || n_j || Salt_{priv}(n_i)).
\end{eqnarray}
Verifiability of the outbound connections allows nodes -- up to some extent -- to verify (or test) if the requesting nodes behave according to the protocol. We call such a test an eligibility test, described in Section \ref{sec:eligibility}.

Node $n_i$ will accept a request from a node $n_j$ if the requesting node $n_{j}$ passes the eligibility test and either:
\begin{itemize}
    \item $n_i$ has less than $k$ inbound connections or
    \item  $S_{in}(n_i,n_j)$ is smaller than the inbound score of one of his accepted peers. In this case, node $n_i$ will drop the connection to its neighbor with the highest score.
\end{itemize}

\subsection{Verifiability and eligibility tests}\label{sec:eligibility}
The eligibility test uses publicly available information to check whether a node follows the protocol, which may give additional protection for the inbound connections. For instance, in the absence of such a test, an adversary can attack inbound connections by sending more requests than it was supposed to send. Eligibility tests may include additional information as for example age of the node, the outbound scores $S_{out}$, or the number of neighbors of the requesting nodes. In this work we discuss the following test:

\subsubsection*{The $\theta$-test}
Let $\theta \in [0,1]$ and assume that node  $n_{i}$ requests a connection with node $n_{j}$. Then $n_{i}$ passes the $\theta$-test of node $n_{j}$ if $S_{out}(n_{i}, n_{j}) \leq \theta$.

The $\theta$-test reduces the strength of an attacker by requiring the outbound score of a requesting node to be no larger than a value $\theta$. Thus, the average number of requests that an adversary controlling $N_A$ nodes can send is $\theta N_A$. Furthermore, to maintain the effectiveness of this threshold  criterion, the value of $\theta$ could be decreased as $N$ increases.  Although a lower $\theta$ value would decrease the proportion of adversary requests, the value should also not be chosen excessively low, since it might also impact honest requests. 

The public verifiability of the outbound connections may increase resistance against eclipse attacks, but also reveals pieces of information on network topology, facilitating attacks. We refer to Section \ref{sec:discussion} for a discussion on this tradeoff.

\subsection{Salt generation}
Nodes need to have a public and private salt, which should be updated periodically after some time $T$. 
The public salt must satisfy the following requirements:
\begin{itemize}
    \item[S1] Future salts must be unguessable. Otherwise, an attacker could mine node ids with small outbound scores for a targeted node.  This offers protection for the outbound connections.
    \item[S2] Salts must not be arbitrarily chosen. Otherwise, an attacker could mine a public salt such that a victim has a low outbound score from the point-of-view of the attacker, thus circumventing the $\theta$-test.  
\end{itemize}
We propose to set the public salt using hash chains, while the private salt can be randomly generated on the fly. Every new node creates a hash chain and publishes its last element as their initial salt. 
After a time interval $T$, the nodes update their public salt by changing it to the preceding value in the hash chain. Upon request of a connection, the requesting node can inform the requested node about the number of salt updates since the last communication. The requested node can then check, by computing a small series of hashes, if the requesting node has followed the protocol.

\subsection{Salt updates and network reorganization}
Nodes should periodically update both of their salts, which leads to protection from brute force attacks, prevents the network to be stuck in states where nodes do not find new peers, and mitigates attacks that use knowledge on the topology. The salt updates should happen asynchronously to limit the negative effects of too heavy network re-organizations. For instance, a synchronous update may lead to all neighbors of a given node dropping the connection at the same time, making this node susceptible to eclipse attacks.

After a public salt update, the node starts looking for new outbound neighbors. It does not immediately drop its previous connections but will replace the existing connections as soon as it finds a new outbound neighbor with a small score that is willing to peer. After a private salt update, a node starts accepting incoming connections that have smaller scores than the current inbound neighbors.

The salt updates will regularly cause a certain reorganization of the network. This  enables new nodes to join more easily the network, since there will always be nodes that recently updated their salt, which are more likely to accept new requests. Since the expiry of salts is not synchronized, the reorganization will affect a small fraction of the network at a given point in time, and negative effects are limited. 

Typically, the evolution of the network topology depends on the average time between salt changes. If the frequency of salt changes is low, the network can get stuck in situations where not every node has the maximal number of peers. Moreover, stable network situations facilitate attacks that are based on knowledge of the network topology. On the other hand, a high frequency of salt updates can lead to situations of constant network reorganization and instability. 
 
\begin{figure}[h]
\vspace{-0.3cm}
\begin{minipage}{.47\textwidth}
    \includegraphics[width=1.0\textwidth,trim={0 .1cm 0 1cm},clip]{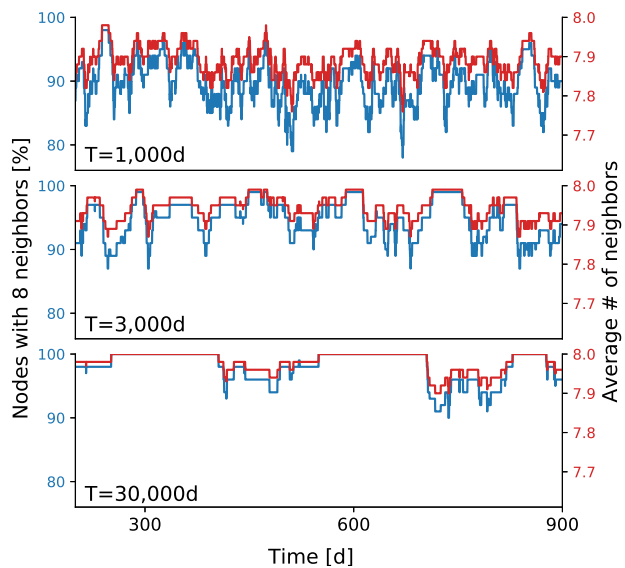}
    \caption{\small Nodes with 8 neighbors (blue) and average number of neighbors (red) with time, for $N=100$ and some values of $T$.}
\label{fig:peer}    
\end{minipage}\hfill
\end{figure}

We illustrate the influence of the salt update frequency on the network topology in Fig. \ref{fig:peer}, for a network of $N$ nodes and $k=4$. We measure the salt update interval $T$ as multiples of the delay $d$ between queries of a node.  We refer to the simulation's open source code \cite{simulator} for more details.

\subsection{Framework}
The above proposal can be seen as one framework among many permutations. The main and essential ingredients of the neighbor selection are outbound and inbound scores. These can be obtained by other means, e.g., XOR-distances, or involve features of additional Sybil protections, e.g., staked funds or the age of a node. The scores can be private or public and also be chosen symmetrically, or the number of incoming and outgoing connections may differ.

\section{Discussion}\label{sec:discussion}
Many aspects of a DLT that can influence the design of an autopeering protocol. Not every aspect can be measured quantitatively and some tradeoffs may only be discussed qualitatively. For instance, every P2P is vulnerable to Sybil attacks. The existence of an autopeering itself may increase the vulnerabilities to Sybil attacks since it facilitates the creation of identities. On the other hand, a P2P that relies on manual peering is not scalable and can lead to network topologies that are suboptimal for the dissemination of information, are easily attackable, or may enable attacks based on social engineering. 

When the benefits of autopeering outweigh possible risks, the autopeering needs a good Sybil protection. This fact is well-known and resistance to Sybil attacks can be strengthened on the neighbor selection layer by increasing the number of connections or filtering the nodes by their age. 
The salt-based autopeering introduced in this paper offers an additional protection through eligibility tests. Note that the eligibility test may take other parameters into account, such as the age of the node, response time in ping-pongs, amount of tokens staked, or more general reputation models. 


The above points are only some aspects of design decisions. A complete analysis of the autopeering  can only be done by including peer selection  and information dissemination in the discussion. Moreover, many of the network layer aspects are far from being well understood and the design of the autopeering must also take into account the specific requirements of the underlying DLT.

\section{Conclusion}
We propose a salt-based neighbor selection protocol that allows local verifiability. We show that this verifiability can increase the protection against eclipse attacks and discuss how negative effects on the topology hiding can be mitigated.  The resulting network will be self-regulating (due to the scores and the eligibility tests) and self-organizing (due to the salted scores). At the same time, it will be dynamic (due to constant reorganizations of parts of the network), allowing new nodes to join the network and mitigating attacks based on the knowledge of the network topology.







\bibliographystyle{IEEEtran}
\bibliography{bibliography}
%

\end{document}